# Empirical examination of the stability of expectations – Augmented Phillips Curve for developing and developed countries[1]

**Yhlas SOVBETOV**
Istanbul University, Turkey
yhlas.sovbetov@ogr.iu.edu.tr
**Muhittin KAPLAN**
Istanbul University, Turkey
muhittin.kaplan@istanbul.edu.tr

**Abstract.** *The empirical literature provides mixed results on the relationship between inflation and unemployment; therefore, there is no consensus on validity and stability of the Phillips Curve. It also seems to be closely related with country-specific factors and the examination time periods. Considering the importance of this trade-off for policy-makers, this study aims to examine validity and stability of expectations-augmented Phillips Curve across 41 countries focusing on three different time periods between 1980 and 2016. The study documents several findings both in country-specific and in panel estimation analysis. First, we find that forward-looking characteristic of inflation picks up weight after 1990's which indicates that inflation became more sensitive to the expected prices. Second, we observe that inflation in developed markets is more forward-looking comparing to emerging and frontier markets. This indicates that developed markets dear forward-looking price expectations more than other markets. Third, we find that that both forward- and backward-looking Phillips Curve fails to work in Brazil, Greece, Indonesia, Mexico, South Africa, Romania, and Turkey. We address it to their long history of high and volatile inflation.*

## 1. Introduction

Since its discovery, the so-called Phillips curve that shows the negative relationship between the rate of wage growth and unemployment has played an important role in macroeconomic modelling. The existence of such a link means that nominal indicators, such as prices, are associated, at least in the short term, with real variables, for example, GDP or unemployment. This relationship has gathered a wide focus on academic sphere (Brissimis and Magginas, 2008; Henzel and Wollmershuser, 2008; Shapiro, 2008; Russel and Banerjee, 2008; Agenor and Bayraktar, 2010; Del Boca et al., 2010; Fendel et al., 2011; Abbas and Sgro, 2011; Benigno and Ricci, 2011; Hall, 2011; Mazumder, 2011; Vasicek, 2011; Rulke, 2012; Ball and Mazumder, 2015; Coibion and Gorodnichenko, 2015), particularly among Keynesian economists. They argue that it offers a favourable control for policymakers in increasing output and lowering unemployment for the sake of higher inflation by expanding aggregate demand. However, opposing views believe that this kind of trade-off might exist only in the short-run as inflationary expectations are corrected according to changes in actual inflation which shifts the short-term Phillips curve and makes the long-term Phillips curve vertical at the point of natural unemployment.

Notwithstanding these criticisms, the Phillips curve has become one of the most prominent foundations for macroeconomics and it has been frequently subjected to empirical analyses in various countries over different time horizons (Brissimis and Magginas, 2008; Russel and Banerjee, 2008; Agenor and Bayraktar, 2010; Del Boca et al., 2010; Fendel et al., 2011; Abbas and Sgro, 2011; Vasicek, 2011; Rulke, 2012). Although many studies (Fitzenberger et al., 2007; Brissimis and Magginas, 2008; Henzel and Wollmershuser, 2008; Sian, 2010; Fendel et al., 2011; Rulke, 2012; Ball and Mazumder, 2015) document existence of Phillips relationship, few exercises have rendered both capability and stability of the Phillips curve to be questioned (Russel and Banerjee, 2008; Paul, 2009; Ojapinwa and Esan, 2013; Nub, 2013; Simionescu, 2014; Coibion and Gorodnichenko, 2015).

In turn, this study offers new evidences on the Phillips curve over 41 different countries with an up-to-date data covering 1980-2016 periods. It tests the validity and stability of the Phillip relation with expectation-augmented model by exploring its implications for the behaviour of inflation and output in developed, emerging, and frontier markets. The study presents new evidence for country-specific Phillips Curve application and for its stability showing how expectations-augmented Phillips Curve evolves over time.

The rest of the paper is structured as follows. Next section briefly reviews formulation of the Phillips curve and integration of expectations into the model. Third section discusses about empirical literature. Forth section describes data and specifies the methodology for this study. Fifth section presents findings and interprets them thoroughly. The final section concludes.

## 2. Literature review of the Phillips Curve

The study of William Phillips (1958) on wage inflation and unemployment in the UK over 1861-1957 has shifted the mainstream of macroeconomics. Phillips in his study observes an inverse correlation: when unemployment goes up, wages start to decrease slowly, and when unemployment drops to low levels, wages tend to rise rapidly. He believes that this happens due to looseness (tightness) of labour market during high (lower) unemployment rates. He formulates the model as:

$$Y + a = bX^{c} + \varepsilon_t \quad (1.1)$$

where $Y$ and $X$ are nominal wage inflation and unemployment rate respectively. He converts the equation into logarithmic form as:

$$log(Y + a) = \log(b) + c * log(X) + \varepsilon_t \quad (1.2)$$

where $b$ is an intercept, and $a$ is a constant added to inflation rate to eliminate errors emerged from negative values in logarithm.

Since Phillips (1958), a sizeable supporting literature has been written on this topic. For instance, Samuelson and Solow (1960) replicate Phillip's study for United States assuming that prices are set by a certain mark-up, and find more consistent results. Although Phillips Curve (PC) was widely accepted and used by policy makers who wanted to benefit from the empirical trade-off in early 1960s, it has received a lot of criticism. Indeed, these comments helped the originally portrayed model to evolve over time reflecting the theoretical developments of the last half-century. For instance, Phelps (1967) and Friedman (1968) criticize the PC arguing that the presented trade-off might occur only in short-run. Moreover, equilibrium in the labour market is determined by the real wage, so PC does not hold as it accounts only money wages (Akerlof, 2007). Besides, Phelps and Friedman believe that structural and frictional unemployment is never-ending, so it is the "natural rate" of unemployment at which inflation rate remains stable (latter it became known as non-accelerating inflation rate of unemployment, NAIRU). If government persistently generates inflationary policies to reduce unemployment below this natural rate, there will be a temporary trade-off (Samuelson and Solow, 1960). As individuals adjust their current expectations accordingly, the trade-off will disappear and unemployment rate will revert back to the natural level. In this respect, the curve will be vertical in long-run which means that unemployment becomes irrelevant to the level of inflation. Thus, a general equation for expectation-augmented Phillips Curve (EAPC)[(2)] model can be expressed as follows.

$$\pi_t = \pi_t^{e} + (\mu + z) - \alpha(u_t - u^{*}) + \varepsilon_t \quad (1.3)$$

where $\pi_t$ is an inflation at time $t$; $\pi^{e}_{t}$ is expected inflation at time $t$; $\mu$ is a mark-up factor that used in determination of market prices; $z$ is a shift factor of labour market; $u_t$ and $u^{*}$ are current (at time $t$) and NAIRU unemployment rates; and $\alpha$ is a parameter of unemployment.

The original Phillip's equation becomes a special case of equation 1.3 where $\pi^{e}$ is zero. In other words, the original Phillips model does not incorporate the behaviour of economic agents into the model. The $\pi^e$ refers to two types of expectations: backward- and forward-looking. In pre-Lucas period, the policy-makers mainly designed their policies assuming that economic agents update their expectations based on the past years' inflation. Thus, they are backward-looking and have adaptive expectations. In such traditional case, $\pi^e_t$ equates to $\zeta\pi_{t-1}$, so equation becomes as $\Delta\pi_t = \frac{\mu+z}{1-\zeta} - \left(\frac{\alpha}{1-\zeta}\right)(u_t - u^*)$. The key idea is that unemployment has an effect on change in rate of inflation, but not on the level of the rate of inflation as it was in original case.

However, Lucas (1976) asserts that incorrect specification of EAPC might generate misleading results especially when it is used for policy evaluation. The Lucas critique highlights that economic agents are not "myopic", instead, they have perfect foresight (forward-looking behaviours) so their expectations are pure rational. They account not only recent information, but all available information, and adjust their expectation instantaneously, so that short-run trade-off between inflation and unemployment does not occur. Here, $\pi^e_t$ is not equal to $\pi_{t-1}$ anymore, instead, it equates to $E_t(\pi_{t+1})$.

Later, Roberts (1995) argues that long-run Phillips Curve might diverge far from being vertical when presence of time discounting and time-contingent nominal contracts (rigidity) are considered. Many Keynesian economists support this view (Fuhrer and Moore, 1995; Gali and Gertler, 1999; Mankiw and Reis, 2002). Thus, focus of academic researches shifted from Neo-Classical synthesis onto New Keynesian synthesis of the Phillips Curve. However, there is no still wide consensus on validity or on demise of EAPC. And studies with more recent date on this topic are lacking in the literature.

In this respect, this study attempts to offer new empirical evidences for both backward- and forward-looking expectation integrated EAPC models over 41 countries since 1980. However, finding a proper proxy for forward-looking expectation term has been always challenging in the related literature. Basically, the expected inflation is a non-observable long-run inflation from which cyclical (temporary) component is stripped off and only more permanent component is left. In literature, Hodrick-Prescott (HP), linear or band-pass filters are frequently utilized for this stripping process to obtain expected inflation (Del Boca et al., 2010). Alternatively, one might assert that the expected inflation should be equal to central bank's long-run inflation target. However, empirical literature shows that both forward- (rational) and backward-looking (adaptive) components play a role in determining how agents set their expectations. Plus, changes in central bank's inflation target and imperfections in monetary policies make stripping temporary and permanent components of inflation more difficult. Thus, there is no consensus in the literature. In addition, instruments such as past years inflation, forecast reports, and surveyed expected inflation are also used in retrieving the expected inflation. Ball and Mazumder (2015) assume that majority in the market anchors their inflation expectations on Federal Reserve's target. Using this proxy, they successfully explain behaviour of inflation after 2000 including the 2008 Great Recession period. Henzel and Wollmershuser (2008) examine Phillips curve by using Survey of Professional Forecasters (SPF) and Ifo World survey forecasts of inflation data and also by employing two-stage least squares to account for potential regressor endogeneity. They document that the Phillips curve performs relatively better when survey forecasts of inflation are utilized. Brissimis and Magginas (2008) also use SPF and Greenbook forecast of inflation data and find that Phillips trade-off appears more significant when survey forecast of inflation is considered rather than the conventional method of using actual known data.

Fendel et al. (2011) use several inflation forecasting polls and they test both traditional and expectation-augmented version of Phillips Curve in G7 countries over October 1989 to December 2007. They observe a significant trade-off between inflation and unemployment thorough out all G7 countries except Italy where traditional Phillips Curve fails to work. However, when they pursue the same analysis with expectation-augmented Phillips Curve they find out that it properly works for all G7 countries where the strongest trade-off appears in Japan and in the US. Rulke (2012) also finds similar trade-off with expectation-augmented Phillips Curve thorough out six Asian-Pacific countries. He observes that the trade-off magnitude is remarkable larger in Japan and South Korea.

Russell and Banerjee (2008) study NAIRU Phillips curve under non-stationarity conditions in the series. They observe a positive relationship between inflation and unemployment rate in short run for the United States, so they conclude that the Phillips Curve does not work. Paul (2009) also fails to document the existence of an empirical Phillips curve in India. He states that the relationship is often evasive or absent in less-developed economies. He underlines that liberalization-policy of the early 1990s and supply shocks such as droughts, oil prices were chief reasons behind the absence of Phillips Curve in India. He shows that a short-run Phillips curve works when adjustments for these shocks are taken into account.

On the other hand, Ojapinwa and Esan (2013) and Simionescu (2014) find weak Phillips relationships for Nigeria and for Romania in short-run respectively, but they disappear in the long-run as inflation and unemployment moves together positively. Sian (2010) provides evidences for validity of Phillips curve in newly industrialized countries in both short- and long-run. He also finds that the trade-off is fairly stable in Mexico and Turkey, whereas it is fragile in Brazil, South Africa, and Thailand.

Fitzenberger et al. (2007) examine NAIRU Phillips curve in Germany and find overall stable relationship where the NAIRU compatible with 2% inflation is associated with 7% unemployment rate. Later, Nub (2013) explores the Phillips trade-off in Germany with an updated data over the period from 1970 to 2012. Unlike Fitzenberger et al. (2007), he fails to detect a significant negative short-run trade-off which indicates that the Phillips Curve is not a reliable relationship for designing policies.

## 3. Data and methodology

The paper empirically tests the validity and the stability of backward-looking and pure forward-looking EAPC over the period1980:Q1-2016:Q1 across the total of 41 developing and developed countries (Appendix A) using eq. 1.3.

$$\pi_t = \beta_0 + \beta_1 \pi_t^e + \beta_2 (u_t - u^*) + \varepsilon_t \quad (1.3)$$

where $\pi_t$ and $\pi_t^e$ are proxied by first differences of logarithm of *CPI* and expected *CPI inflation* over one year respectively. In eq. 1.3 if $\pi_t^e = \pi_{t-1}$, then the model converts to traditional PC that has backward-looking specification. On the other hand, $U_t$, and $U^*$ are proxied by unemployment rate and *NAIRU* respectively in logarithmic form. Data for these variables are obtained from Thomson Reuters Eikon Datastream, and we added a fixed constant term to all series to handle negative values during transformation into logarithmic form which only shifts $\beta_0$ up leaving other variables unaffected. The expected inflation data also retrieved from Thomson Reuters Eikon Datastream where the data is formed by their own forecasting survey methodology.

Table 1 provides an overview for main features of the dataset. The first and second columns of the table present current and expected inflation of CPI derived from Thomson Reuters Eikon Datastream. The third column is derived by HP filtering the current inflation (first column) with lambda 1600. Comparing Thomson's forecasted expected inflation, the HP filter derives more accurate estimation. Thus, we present NAIRU figures at fifth column that are obtained by HP filtering methodology.

**Table 1.** *Overview of Mean and Standard Deviation of Variables*

| | Current Inflation | | Expected Inflation | | HP Inflation Trend | | Unemployment Rate | | NAIRU | | Unemployment Gap | |
|---|---|---|---|---|---|---|---|---|---|---|---|---|
| | *Mean* | *Std. Dev.* | *Mean* | *Std. Dev.* | *Mean* | *Std. Dev.* | *Mean* | *Std. Dev.* | *Mean* | *Std. Dev.* | *Mean* | *Std. Dev.* |
| **AG** | 0.0115 | 0.0151 | 0.0197 | 0.0253 | 0.0176 | 0.0283 | 0.1175 | 0.0447 | 0.1169 | 0.0378 | 0.0006 | 0.0191 |
| **AU** | 0.0027 | 0.0023 | 0.0031 | 0.0008 | 0.0028 | 0.0005 | 0.0645 | 0.0171 | 0.0644 | 0.0160 | 0.0001 | 0.0043 |
| **BD** | 0.0019 | 0.0018 | 0.0022 | 0.0010 | 0.0019 | 0.0009 | 0.0910 | 0.0167 | 0.0905 | 0.0151 | 0.0005 | 0.0056 |
| **BG** | 0.0021 | 0.0020 | 0.0022 | 0.0009 | 0.0021 | 0.0005 | 0.0824 | 0.0089 | 0.0817 | 0.0052 | 0.0007 | 0.0054 |
| **BR** | 0.0476 | 0.1053 | 0.0867 | 0.1978 | 0.0464 | 0.0808 | 0.1145 | 0.0298 | 0.1147 | 0.0267 | -0.0002 | 0.0106 |
| **CH** | 0.0045 | 0.0072 | 0.0058 | 0.0055 | 0.0044 | 0.0042 | 0.0368 | 0.0059 | 0.0368 | 0.0057 | 0.0000 | 0.0011 |
| **CL** | 0.0056 | 0.0052 | 0.0061 | 0.0043 | 0.0057 | 0.0040 | 0.0866 | 0.0196 | 0.0865 | 0.0157 | 0.0001 | 0.0100 |
| **CN** | 0.0019 | 0.0018 | 0.0023 | 0.0008 | 0.0020 | 0.0005 | 0.0780 | 0.0136 | 0.0779 | 0.0116 | 0.0001 | 0.0051 |
| **CZ** | 0.0054 | 0.0076 | 0.0068 | 0.0076 | 0.0054 | 0.0045 | 0.0545 | 0.0187 | 0.0545 | 0.0168 | 0.0000 | 0.0061 |
| **DK** | 0.0020 | 0.0019 | 0.0024 | 0.0007 | 0.0020 | 0.0006 | 0.0662 | 0.0261 | 0.0657 | 0.0224 | 0.0005 | 0.0076 |
| **ES** | 0.0029 | 0.0024 | 0.0032 | 0.0016 | 0.0029 | 0.0015 | 0.1754 | 0.0588 | 0.1736 | 0.0523 | 0.0017 | 0.0142 |
| **FN** | 0.0017 | 0.0020 | 0.0022 | 0.0010 | 0.0018 | 0.0007 | 0.1000 | 0.0294 | 0.0983 | 0.0247 | 0.0016 | 0.0095 |
| **FR** | 0.0016 | 0.0017 | 0.0019 | 0.0007 | 0.0016 | 0.0007 | 0.0899 | 0.0100 | 0.0895 | 0.0078 | 0.0004 | 0.0047 |
| **GR** | 0.0048 | 0.0079 | 0.0053 | 0.0049 | 0.0048 | 0.0046 | 0.1466 | 0.0603 | 0.1458 | 0.0556 | 0.0007 | 0.0155 |
| **HN** | 0.0108 | 0.0116 | 0.0114 | 0.0092 | 0.0110 | 0.0075 | 0.0773 | 0.0179 | 0.0773 | 0.0152 | 0.0001 | 0.0052 |
| **ID** | 0.0101 | 0.0134 | 0.0109 | 0.0121 | 0.0101 | 0.0042 | 0.0668 | 0.0191 | 0.0667 | 0.0174 | 0.0001 | 0.0065 |
| **IN** | 0.0077 | 0.0069 | 0.0078 | 0.0025 | 0.0077 | 0.0021 | 0.0824 | 0.0210 | 0.0823 | 0.0209 | 0.0000 | 0.0005 |
| **IR** | 0.0022 | 0.0029 | 0.0027 | 0.0016 | 0.0022 | 0.0014 | 0.0902 | 0.0417 | 0.0897 | 0.0366 | 0.0005 | 0.0105 |
| **IT** | 0.0026 | 0.0019 | 0.0030 | 0.0016 | 0.0027 | 0.0014 | 0.0957 | 0.0188 | 0.0954 | 0.0170 | 0.0003 | 0.0050 |
| **JP** | 0.0003 | 0.0025 | 0.0006 | 0.0010 | 0.0003 | 0.0007 | 0.0414 | 0.0080 | 0.0413 | 0.0070 | 0.0001 | 0.0028 |
| **KO** | 0.0037 | 0.0034 | 0.0049 | 0.0022 | 0.0038 | 0.0016 | 0.0359 | 0.0121 | 0.0359 | 0.0062 | 0.0000 | 0.0087 |
| **MX** | 0.0095 | 0.0103 | 0.0105 | 0.0096 | 0.0096 | 0.0067 | 0.0395 | 0.0106 | 0.0392 | 0.0076 | 0.0003 | 0.0061 |
| **MY** | 0.0030 | 0.0029 | 0.0042 | 0.0015 | 0.0029 | 0.0007 | 0.0325 | 0.0043 | 0.0324 | 0.0022 | 0.0001 | 0.0033 |
| **NL** | 0.0022 | 0.0021 | 0.0024 | 0.0008 | 0.0022 | 0.0006 | 0.0627 | 0.0165 | 0.0620 | 0.0123 | 0.0007 | 0.0074 |
| **NW** | 0.0022 | 0.0022 | 0.0024 | 0.0007 | 0.0022 | 0.0003 | 0.0388 | 0.0088 | 0.0389 | 0.0068 | 0.0000 | 0.0039 |
| **OE** | 0.0021 | 0.0026 | 0.0023 | 0.0008 | 0.0021 | 0.0007 | 0.0475 | 0.0062 | 0.0475 | 0.0045 | 0.0001 | 0.0036 |
| **PH** | 0.0058 | 0.0043 | 0.0069 | 0.0030 | 0.0060 | 0.0023 | 0.0889 | 0.0184 | 0.0890 | 0.0161 | -0.0001 | 0.0068 |
| **PO** | 0.0088 | 0.0112 | 0.0118 | 0.0147 | 0.0091 | 0.0111 | 0.1410 | 0.0311 | 0.1406 | 0.0216 | 0.0004 | 0.0145 |
| **PT** | 0.0032 | 0.0028 | 0.0036 | 0.0027 | 0.0032 | 0.0022 | 0.0831 | 0.0359 | 0.0825 | 0.0337 | 0.0007 | 0.0095 |
| **RM** | 0.0337 | 0.0456 | 0.0343 | 0.0449 | 0.0300 | 0.0297 | 0.0736 | 0.0243 | 0.0725 | 0.0193 | 0.0011 | 0.0131 |
| **RS** | 0.0481 | 0.0941 | 0.0761 | 0.1456 | 0.0449 | 0.0577 | 0.0777 | 0.0219 | 0.0777 | 0.0187 | 0.0001 | 0.0080 |
| **SA** | 0.0072 | 0.0043 | 0.0081 | 0.0029 | 0.0072 | 0.0021 | 0.2334 | 0.0324 | 0.2337 | 0.0278 | -0.0004 | 0.0109 |

| | Current Inflation | | Expected Inflation | | HP Inflation Trend | | Unemployment Rate | | NAIRU | | Unemployment Gap | |
|---|---|---|---|---|---|---|---|---|---|---|---|---|
| | *Mean* | *Std. Dev.* | *Mean* | *Std. Dev.* | *Mean* | *Std. Dev.* | *Mean* | *Std. Dev.* | *Mean* | *Std. Dev.* | *Mean* | *Std. Dev.* |
| **SD** | 0.0015 | 0.0028 | 0.0022 | 0.0015 | 0.0016 | 0.0014 | 0.0825 | 0.0191 | 0.0812 | 0.0145 | 0.0013 | 0.0085 |
| **SP** | 0.0019 | 0.0028 | 0.0026 | 0.0012 | 0.0019 | 0.0013 | 0.0238 | 0.0074 | 0.0236 | 0.0059 | 0.0002 | 0.0043 |
| **SW** | 0.0009 | 0.0026 | 0.0014 | 0.0013 | 0.0010 | 0.0012 | 0.0342 | 0.0090 | 0.0335 | 0.0049 | 0.0006 | 0.0056 |
| **TH** | 0.0033 | 0.0042 | 0.0044 | 0.0023 | 0.0033 | 0.0016 | 0.0181 | 0.0113 | 0.0181 | 0.0085 | 0.0000 | 0.0063 |
| **TK** | 0.0321 | 0.0292 | 0.0380 | 0.0319 | 0.0322 | 0.0251 | 0.0850 | 0.0190 | 0.0848 | 0.0158 | 0.0002 | 0.0091 |
| **TW** | 0.0016 | 0.0041 | 0.0024 | 0.0014 | 0.0016 | 0.0013 | 0.0372 | 0.0119 | 0.0373 | 0.0103 | 0.0000 | 0.0047 |
| **UK** | 0.0024 | 0.0029 | 0.0031 | 0.0019 | 0.0024 | 0.0012 | 0.0440 | 0.0204 | 0.0435 | 0.0177 | 0.0005 | 0.0049 |
| **US** | 0.0025 | 0.0021 | 0.0028 | 0.0007 | 0.0025 | 0.0007 | 0.0601 | 0.0161 | 0.0602 | 0.0122 | -0.0001 | 0.0072 |
| **VE** | 0.0354 | 0.0265 | 0.0408 | 0.0279 | 0.0358 | 0.0199 | 0.1059 | 0.0335 | 0.1063 | 0.0278 | -0.0004 | 0.0143 |

## 4. Results and discussion

Table 2 presents results of OLS estimation of backward-looking (traditional) and forward-looking EAPC for 41 countries individually in different time horizons. We categorize countries according their S&P market classification, and show EAPC estimations after 1980's, 1990's, and 2000's separately in order to make clear comparisons. The results shed light on an overall validity of the traditional backward-looking EAPC during early 1980's where afterwards it loses significance in majority of the sampled countries. Especially, we detect that the PC was valid for all developed countries except Belgium and Singapore during 1980-1990, whereas its behaviour was ambiguous for emerging and frontier markets. More specifically, the results indicate that PC fails to hold in all Latin American countries – *Chile and Mexico in emerging market, and Argentina and Venezuela in frontier market classification*. This result might be due to LatAm sovereign debt crisis during 1980's. Equally, we detect that traditional PC also fails in Greece and Turkey due to the same reason, i.e. unstable economic environment. These two countries have experience about 20 quarters of recessions just during 1980-1990 (Appendix A).

On the other hand, forward-looking EAPC performs better comparing to traditional one as it is functional for almost all developed countries during 1980-2016, except Singapore who also had failed in the previous (traditional PC) analysis as well. Among emerging and frontier markets, the only improvement occurs in Thailand in which PC becomes valid with pure forward-looking specification. Moreover, notice that the model continues its functionality in developed markets, even, when we restrict the analysis period to 1990-2016 where Thai EAPC gains its validity, but Austrian EAPC loses it. Although, we fail to detect substantial changes for emerging and frontier markets, it is clearly visible that the PC with pure forward-looking specification tend to perform better than the one with backward-looking specification during 1990-2016 comparing to unrestricted period. Apparently, few Latin American countries like Argentina, Chile, and Venezuela generate plausible results regarding solidness of EAPC after overcoming the sovereign debt crisis hurricane in mid 1980's.

**Table 2.** *Backward - and Forward - Looking PC over Different Time Horizons*

| | Periods | After 80's 1980:Q1-2016:Q1 | | | | After 90's 1990:Q1-2016:Q1 | | | | After 00's 2000:Q1-2016:Q1 | | | |
|---|---|---|---|---|---|---|---|---|---|---|---|---|---|
| | | BACKWARD-LOOKING | | FORWARD-LOOKING | | BACKWARD-LOOKING | | FORWARD-LOOKING | | BACKWARD-LOOKING | | FORWARD-LOOKING | |
| M | Variables | $\pi_{t-1}$ | $U_{GAP}$ | $\pi^e_{t+1}$ | $U_{GAP}$ | $\pi_{t-1}$ | $U_{GAP}$ | $\pi^e_{t+1}$ | $U_{GAP}$ | $\pi_{t-1}$ | $U_{GAP}$ | $\pi^e_{t+1}$ | $U_{GAP}$ |
| D | Australia | 0.5334***<br>(0.1134) | -0.1255**<br>(0.0522) | 0.5646***<br>(0.1053) | -0.0754**<br>(0.0312) | 0.3549**<br>(0.1455) | -0.1565**<br>(0.0644) | 0.5810**<br>(0.2500) | -0.1009**<br>(0.0501) | 0.3073**<br>(0.1481) | -0.2259*<br>(0.1352) | 0.6724***<br>(0.1777) | -0.0559*<br>(0.0304) |
| D | Austria | 0.4862***<br>(0.0802) | -0.1337***<br>(0.0416) | 0.5979***<br>(0.0914) | -0.1064***<br>(0.0385) | 0.4940***<br>(0.0953) | -0.0552<br>(0.0375) | 0.7481***<br>(0.1648) | -0.0342<br>(0.0383) | 0.5695***<br>(0.1371) | 0.0119<br>(0.0642) | 1.2965***<br>(0.1991) | -0.0084<br>(0.0687) |
| D | Belgium | 0.5199***<br>(0.0805) | -0.0414<br>(0.0346) | 0.8557***<br>(0.0631) | -0.0144**<br>(0.0075) | 0.3869***<br>(0.1184) | -0.0286<br>(0.0438) | 0.5775***<br>(0.1242) | -0.0531**<br>(0.0247) | 0.2257*<br>(0.1327) | -0.3054***<br>(0.1015) | 0.5581***<br>(0.1952) | -0.2605***<br>(0.1043) |
| D | Canada | 0.5225***<br>(0.1100) | -0.1082**<br>(0.0434) | 0.5547***<br>(0.0708) | -0.0630**<br>(0.0251) | 0.1358*<br>(0.0755) | -0.1703***<br>(0.0592) | 0.5752***<br>(0.1165) | -0.1592***<br>(0.0528) | 0.0193<br>(0.0705) | -0.1435**<br>(0.0633) | -0.9886<br>(0.6668) | -0.2331**<br>(0.1001) |
| D | Denmark | 0.5968***<br>(0.1038) | -0.0239<br>(0.0238) | 0.4660***<br>(0.1076) | -0.0387**<br>(0.0166) | -0.0066<br>(0.0904) | -0.0063<br>(0.0205) | 0.5225***<br>(0.1327) | -0.0378**<br>(0.0152) | 0.0660<br>(0.0887) | -0.0529<br>(0.0644) | 0.6444***<br>(0.2059) | -0.0488*<br>(0.0259) |
| D | Finland | 0.5877***<br>(0.0761) | -0.0343*<br>(0.0213) | 0.5026***<br>(0.0685) | -0.0646***<br>(0.0165) | 0.3110***<br>(0.0830) | -0.0758**<br>(0.0306) | 0.5308***<br>(0.1423) | -0.0770***<br>(0.0187) | 0.2182*<br>(0.1308) | -0.2448***<br>(0.0837) | 0.6087***<br>(0.1919) | -0.1796**<br>(0.0805) |
| D | France | 0.5059***<br>(0.0486) | -0.0472*<br>(0.0272) | 0.5843***<br>(0.0329) | -0.0631**<br>(0.0319) | 0.4326***<br>(0.1401) | -0.0714<br>(0.0601) | 0.5621***<br>(0.1255) | -0.0615**<br>(0.0302) | 0.3312*<br>(0.1815) | -0.0811<br>(0.0837) | 0.5987**<br>(0.2553) | -0.0524*<br>(0.0312) |
| D | Germany | 0.4059***<br>(0.1052) | -0.1024**<br>(0.0510) | 0.5651***<br>(0.0980) | -0.0884**<br>(0.0427) | 0.1482<br>(0.1024) | -0.1074**<br>(0.0513) | 0.5988***<br>(0.1665) | -0.0469**<br>(0.0195) | 0.1146<br>(0.1301) | -0.0171<br>(0.0704) | 0.6890**<br>(0.2918) | -0.0239*<br>(0.0128) |
| D | Ireland | 0.4963***<br>(0.0694) | -0.0539**<br>(0.0250) | 0.5054***<br>(0.0831) | -0.0570*<br>(0.0302) | 0.3609***<br>(0.1317) | -0.0613***<br>(0.0231) | 0.5623***<br>(0.1184) | -0.0318**<br>(0.0150) | 0.4078***<br>(0.1276) | -0.0712<br>(0.0609) | 0.9864***<br>(0.2020) | -0.0300<br>(0.0202) |
| D | Italy | 0.8812***<br>(0.0428) | -0.0353***<br>(0.0115) | 0.9194***<br>(0.0412) | -0.0498***<br>(0.0138) | 0.6800***<br>(0.0812) | -0.0384***<br>(0.0143) | 0.8459***<br>(0.0751) | -0.0456***<br>(0.0144) | 0.8259***<br>(0.1441) | -0.0405<br>(0.0381) | 0.8955***<br>(0.3164) | -0.0437<br>(0.0354) |
| D | Japan | 0.3674***<br>(0.1391) | -0.3150**<br>(0.1372) | 0.5398***<br>(0.1479) | -0.2361**<br>(0.1127) | 0.3115**<br>(0.1507) | -0.3167**<br>(0.1441) | 0.7525***<br>(0.2688) | -0.1601*<br>(0.0939) | 0.4739***<br>(0.1682) | -0.1065<br>(0.1139) | 0.7222*<br>(0.4503) | -0.1026<br>(0.1233) |
| D | Netherlands | 0.3765***<br>(0.1165) | -0.0419*<br>(0.0234) | 0.5774***<br>(0.1778) | -0.0289*<br>(0.0153) | 0.2675**<br>(0.1216) | -0.0404**<br>(0.0208) | 0.5482***<br>(0.1075) | -0.0252***<br>(0.0052) | 0.2809*<br>(0.1547) | -0.0320<br>(0.0285) | 0.9985***<br>(0.3602) | 0.0106<br>(0.0420) |
| D | Norway | 0.4784***<br>(0.0952) | -0.2049**<br>(0.0883) | 0.4938***<br>(0.0738) | -0.1091*<br>(0.0645) | -0.1165<br>(0.0733) | -0.0194<br>(0.0895) | 0.5166***<br>(0.1580) | -0.0544**<br>(0.0241) | -0.1538<br>(0.1411) | -0.0987<br>(0.1422) | 0.6824<br>(0.6495) | -0.0776<br>(0.1493) |
| D | Portugal | 0.4846***<br>(0.0685) | -0.0558*<br>(0.0303) | 0.5852***<br>(0.0662) | -0.0398*<br>(0.0227) | 0.4657***<br>(0.1064) | -0.0974***<br>(0.0341) | 0.6213***<br>(0.1256) | -0.0325*<br>(0.0187) | 0.0341<br>(0.1741) | -0.0697<br>(0.0515) | 0.9466***<br>(0.2009) | 0.0016<br>(0.0291) |
| D | Singapore | 0.5019***<br>(0.1018) | -0.0177<br>(0.0156) | 0.6295***<br>(0.1052) | -0.0117<br>(0.0237) | 0.3864***<br>(0.1240) | -0.1305*<br>(0.0783) | 0.6556***<br>(0.1958) | -0.2202***<br>(0.0910) | 0.3156**<br>(0.1444) | -0.2107*<br>(0.1147) | 0.7046**<br>(0.3228) | -0.2850**<br>(0.1270) |
| D | South Korea | 0.4731***<br>(0.1297) | -0.1272**<br>(0.0649) | 0.5254***<br>(0.1642) | -0.0954*<br>(0.0552) | 0.4269***<br>(0.1333) | -0.2877*<br>(0.1537) | 0.6705***<br>(0.1367) | -0.2808*<br>(0.1584) | 0.2929**<br>(0.1232) | 0.0954<br>(0.1624) | 0.6214**<br>(0.3102) | 0.2029<br>(0.1728) |
| D | Spain | 0.5165***<br>(0.0912) | -0.0519***<br>(0.0178) | 0.6279***<br>(0.0663) | -0.0475*<br>(0.0263) | 0.3069***<br>(0.1136) | -0.0393*<br>(0.0234) | 0.4838***<br>(0.1487) | -0.0403*<br>(0.0232) | 0.2533*<br>(0.1415) | -0.0559<br>(0.0345) | -0.1284<br>(0.6760) | -0.0708<br>(0.0542) |

| | **Periods** | **After 80's 1980:Q1-2016:Q1** | | | | **After 90's 1990:Q1-2016:Q1** | | | | **After 00's 2000:Q1-2016:Q1** | | | |
|---|---|---|---|---|---|---|---|---|---|---|---|---|---|
| | | **BACKWARD-LOOKING** | | **FORWARD-LOOKING** | | **BACKWARD-LOOKING** | | **FORWARD-LOOKING** | | **BACKWARD-LOOKING** | | **FORWARD-LOOKING** | |
| **M** | **Variables** | $\pi_{t-1}$ | $U_{GAP}$ | $\pi^e_{t+1}$ | $U_{GAP}$ | $\pi_{t-1}$ | $U_{GAP}$ | $\pi^e_{t+1}$ | $U_{GAP}$ | $\pi_{t-1}$ | $U_{GAP}$ | $\pi^e_{t+1}$ | $U_{GAP}$ |
| D | Sweden | 0.5188***<br>(0.0828) | -0.0812**<br>(0.0388) | 0.7693***<br>(0.0747) | -0.0606*<br>(0.0322) | 0.2679***<br>(0.1042) | -0.0994*<br>(0.05222) | 0.8862***<br>(0.1248) | -0.0691*<br>(0.0402) | 0.3090**<br>(0.1511) | -0.1409*<br>(0.0846) | 0.7011*<br>(0.4157) | -0.1151<br>(0.1113) |
| D | Switzerland | 0.7105***<br>(0.0532) | -0.0897*<br>(0.0455) | 0.6547***<br>(0.0759) | -0.1393*<br>(0.0779) | 0.7462***<br>(0.0699) | -0.0449<br>(0.0414) | 0.9258***<br>(0.0763) | -0.0556*<br>(0.0329) | 0.5026***<br>(0.1007) | -0.0027<br>(0.0657) | 0.5851*<br>(0.03536) | -0.0154<br>(0.1020) |
| D | United Kingdom | 0.4518***<br>(0.0732) | -0.0498**<br>(0.0254) | 0.6460***<br>(0.0622) | -0.1464**<br>(0.0705) | 0.3655***<br>(0.1109) | -0.0758*<br>(0.0445) | 0.6492***<br>(0.1206) | -0.1533**<br>(0.0782) | 0.4828***<br>(0.1098) | 0.1011<br>(0.0807) | 0.1350<br>(0.1523) | 0.0981<br>(0.0725) |
| D | United States | 0.4065***<br>(0.1269) | -0.0944***<br>(0.0302) | 0.7313***<br>(0.1243) | -0.0919***<br>(0.0344) | 0.1695***<br>(0.0708) | -0.0826***<br>(0.0292) | 0.7489**<br>(0.3622) | -0.0772**<br>(0.0360) | 0.1195*<br>(0.0707) | -0.0779**<br>(0.0315) | 0.1043<br>(0.8176) | -0.0540<br>(0.0459) |
| E | Brazil | - | - | - | - | 0.9041***<br>(0.0874) | 0.2974<br>(0.4276) | 0.5077***<br>(0.0401) | 0.8294<br>(0.5782) | 0.5023***<br>(0.1597) | -0.0132<br>(0.0431) | 1.2063***<br>(0.4122) | -0.0356<br>(0.0462) |
| E | Chile | 0.7355***<br>(0.0798) | -0.0664<br>(0.0429) | 0.9106***<br>(0.0451) | -0.0523<br>(0.0533) | 0.6565***<br>(0.0858) | -0.0577<br>(0.0395) | 0.9049***<br>(0.0465) | -0.0803*<br>(0.0492) | 0.3150***<br>(0.1135) | -0.1243**<br>(0.0670) | 1.1182***<br>(0.3034) | -0.1299*<br>(0.0738) |
| E | China | 0.5746***<br>(0.0654) | -0.0277*<br>(0.0143) | 0.5367***<br>(0.1106) | -0.0364*<br>(0.0203) | 0.5183***<br>(0.0688) | -0.0481*<br>(0.0264) | 0.5221***<br>(0.1527) | -0.0467*<br>(0.0262) | 0.5651***<br>(0.0862) | -0.0013<br>(0.0282) | 0.5594*<br>(0.3184) | -0.0273*<br>(0.0160) |
| E | Czech | - | - | - | - | 0.5396***<br>(0.1171) | -0.0522<br>(0.0739) | 0.4289***<br>(0.1159) | -0.0738*<br>(0.0391) | 0.3115**<br>(0.1384) | -0.0982<br>(0.0707) | 0.2125<br>(0.3554) | -0.1444*<br>(0.0853) |
| E | Greece | 0.9058***<br>(0.0286) | -0.0123<br>(0.0159) | 0.5613***<br>(0.0834) | -0.0163<br>(0.0549) | 0.8898***<br>(0.0427) | -0.0119<br>(0.0133) | 0.8442***<br>(0.0901) | -0.0101<br>(0.0281) | 0.5389***<br>(0.1069) | -0.0187<br>(0.0192) | 1.1177***<br>(0.2270) | -0.0155<br>(0.0235) |
| E | Hungary | - | - | - | - | 0.5463**<br>(0.2335) | -0.1366<br>(0.1958) | 0.4612***<br>(0.0988) | -0.2238<br>(0.1920) | 0.0872<br>(0.1086) | -0.2410<br>(0.2527) | -0.0661<br>(0.9728) | -0.2595<br>(0.3038) |
| E | India | 0.4686***<br>(0.0765) | -0.0620**<br>(0.0277) | 0.4190***<br>(0.0838) | -0.1562***<br>(0.0378) | 0.4215***<br>(0.0822) | -0.1506***<br>(0.0442) | 0.5258***<br>(0.2096) | -0.1326**<br>(0.0576) | 0.4204***<br>(0.1107) | -0.1175**<br>(0.0576) | 0.6434<br>(0.5338) | -0.1399*<br>(0.0837) |
| E | Indonesia | 0.6439***<br>(0.1485) | -0.0712<br>(0.1155) | 0.6189***<br>(0.1462) | -0.0591<br>(0.1317) | 0.6728***<br>(0.1507) | -0.0564<br>(0.1275) | 0.6685***<br>(0.1548) | -0.0282<br>(0.1470) | 0.1517<br>(0.1797) | 0.3154<br>(0.3712) | 0.7554***<br>(0.3056) | 0.2431<br>(0.3622) |
| E | Malaysia | 0.5434***<br>(0.1589) | -0.0426*<br>(0.0226) | 0.4253***<br>(0.1244) | -0.0571***<br>(0.0220) | 0.4674***<br>(0.1233) | -0.0480*<br>(0.0271) | 0.4179***<br>(0.1679) | -0.0692**<br>(0.0348) | 0.1183<br>(0.1812) | -0.0675*<br>(0.03811) | 0.5377<br>(0.6037) | -0.0803*<br>(0.0437) |
| E | Mexico | 0.8778***<br>(0.0841) | 0.2121<br>(0.2052) | 0.9072***<br>(0.0300) | 0.0404<br>(0.1741) | 0.7153***<br>(0.0665) | 0.0551<br>(0.2056) | 0.9184***<br>(0.0918) | -0.0451<br>(0.1438) | 0.1107<br>(0.0851) | -0.1895***<br>(0.0675) | 0.6571***<br>(0.1062) | -0.0751*<br>(0.0414) |
| E | Philippines | 0.6027***<br>(0.0928) | -0.1999***<br>(0.0591) | 0.5921***<br>(0.0784) | -0.0752**<br>(0.0387) | 0.5421***<br>(0.1510) | -0.1175**<br>(0.0567) | 0.6216***<br>(0.0581) | -0.0289**<br>(0.0139) | 0.3249***<br>(0.1375) | -0.0167<br>(0.0471) | 0.5652***<br>(0.2885) | 0.0026<br>(0.0411) |

| M | Periods | After 80's 1980:Q1-2016:Q1 | | | | After 90's 1990:Q1-2016:Q1 | | | | After 00's 2000:Q1-2016:Q1 | | | |
|---|---|---|---|---|---|---|---|---|---|---|---|---|---|
| | | BACKWARD-LOOKING | | FORWARD-LOOKING | | BACKWARD-LOOKING | | FORWARD-LOOKING | | BACKWARD-LOOKING | | FORWARD-LOOKING | |
| | Variables | $\pi_{t-1}$ | $U_{GAP}$ | $\pi^e_{t+1}$ | $U_{GAP}$ | $\pi_{t-1}$ | $U_{GAP}$ | $\pi^e_{t+1}$ | $U_{GAP}$ | $\pi_{t-1}$ | $U_{GAP}$ | $\pi^e_{t+1}$ | $U_{GAP}$ |
| E | Poland | - | - | - | - | 0.6058***<br>(0.1015) | 0.0156<br>(0.0911) | 0.7878***<br>(0.1041) | -0.4411<br>(0.4678) | 0.4155***<br>(0.1325) | -0.0417**<br>(0.0212) | 0.8524***<br>(0.1750) | -0.0422*<br>(0.0241) |
| E | Russia | - | - | - | - | 0.6375***<br>(0.1403) | -1.1854<br>(1.2084) | 0.4140***<br>(0.0537) | -1.4802<br>(1.6148) | 0.5430***<br>(0.1337) | -0.1281***<br>(0.0500) | 0.5714***<br>(0.1829) | -0.2154***<br>(0.0863) |
| E | South Africa | 0.7289***<br>(0.0505) | -0.0008<br>(0.0290) | 0.8722***<br>(0.0699) | 0.0111<br>(0.0309) | 0.6614***<br>(0.0864) | 0.0002<br>(0.0280) | 0.8970***<br>(0.0991) | 0.0103<br>(0.0318) | 0.5934***<br>(0.1149) | -0.0141<br>(0.0319) | 0.9257***<br>(0.3588) | -0.0192<br>(0.0336) |
| E | Taiwan | 0.4072**<br>(0.2127) | -0.2695***<br>(0.0572) | 0.5379***<br>(0.1741) | -0.2232***<br>(0.0713) | -0.1479*<br>(0.0886) | -0.3292***<br>(0.0585) | 0.0354<br>(0.2476) | -0.2636***<br>(0.0978) | 0.0149<br>(0.0754) | -0.1834*<br>(0.0968) | 0.4622<br>(0.5004) | -0.0877<br>(0.1058) |
| E | Thailand | 0.5648***<br>(0.0701) | -0.0496<br>(0.0315) | 0.5958***<br>(0.1171) | -0.0909**<br>(0.0465) | 0.5249***<br>(0.0843) | -0.0778**<br>(0.0385) | 0.4950***<br>(0.1802) | -0.1309**<br>(0.0573) | 0.4337***<br>(0.0942) | -0.0282<br>(0.0556) | 0.2918<br>(0.4648) | -0.0860<br>(0.0839) |
| E | Turkey | 0.7795***<br>(0.0670) | 0.0119<br>(0.1490) | 0.7698***<br>(0.0414) | -0.0148<br>(0.0868) | 0.8442***<br>(0.0713) | -0.0115<br>(0.1217) | 0.8123***<br>(0.0369) | -0.0269<br>(0.0766) | 0.7450***<br>(0.0602) | -0.0748<br>(0.0950) | 0.8669***<br>(0.0884) | -0.0886<br>(0.0718) |
| F | Argentina | 0.6792***<br>(0.1233) | -0.3090<br>(0.6578) | 0.7832***<br>(0.0207) | 0.0697<br>(0.1914) | 0.6264***<br>(0.0981) | -0.1693<br>(0.1579) | 0.7998***<br>(0.0068) | -0.1627*<br>(0.0964) | 0.7913***<br>(0.1041) | -0.40631*<br>(0.2317) | 0.5625***<br>(0.0797) | -0.2628**<br>(0.1333) |
| F | Romania | - | - | - | - | 0.3838*<br>(0.2234) | -0.5001<br>(0.4835) | 0.7679***<br>(0.1476) | 0.4389<br>(0.3580) | 0.8666***<br>(0.0469) | -0.0335<br>(0.0410) | 0.9103***<br>(0.0621) | 0.0077<br>(0.0520) |
| F | Venezuela | 0.8261***<br>(0.1007) | -0.1267<br>(0.0797) | 0.4273***<br>(0.1640) | 0.0918<br>(0.2183) | 0.8708***<br>(0.1027) | -0.1408*<br>(0.0814) | 0.8371***<br>(0.0860) | -0.3243***<br>(0.0821) | 0.9075***<br>(0.1568) | -0.1538<br>(0.1055) | 0.8859***<br>(0.1251) | -0.2921***<br>(0.0809) |

**Notes:** Numbers in the table are coefficient estimates with HAC standard errors in parentheses. The *, **, and *** denote significance at 10%, 5%, and 1% levels respectively. The market classification is in S&P standards. The inflation is denoted as π and proxied by CPI. And $U_{GAP}=U-U^*$.

More interestingly, we observe that coefficients of backward-looking specification decrease as we get closer to the present time, while coefficients of forward-looking specification get dominance. It seems to be valid for almost all countries, especially for developed ones, and indicates that backward-looking specification was a fundamental factor in dynamics of inflation in early 80's which, by the time, is replaced by forward-looking specification. It also explains why inflation today is more prone to market speculations/expectations than it was before 1980's.

Furthermore, when the analysis period is restricted to 2000-2016, we observe that the model has weakened in terms of overall significance almost in all sampled countries. During this period we account massive failures, especially in developed markets which can be addressed to the number of economic crises – *Dotcom crisis (1999-2001), Great Recession (2007-2009), and Eurozone sovereign debt crisis (2009-2014)* – comprised by the analysis time interval. Thus, we conclude that PC with pure forward-looking specification performs better than that with backward-looking one – *which loses significance after 1990's* – until 2000's where afterwards it also collapses.

## 5. Conclusion

This study reports new evidences for validity and stability of EAPC over 41 different countries from developed, emerging, and frontier markets. First, we find that both backward-and forward-looking (BF) EAPC performs better in developed markets comparing to countries from other markets. We address it to well-established, smoothly working, and freely operating structure of economic systems in developed markets.

Second, we observe that forward-looking characteristic of inflation picks up weight after 1990's which indicates that inflation became more sensitive to expected prices. We observe that inflation in developed markets (with nearly 40-60% BF trade-off) is more forward-looking comparing to emerging (50-50% BF trade-off) and frontier (60-40% BF trade-off) markets. This indicates that developed markets dear forward-looking price expectations more than others. Thus, we conclude that inflation is mainly anchored in frontier markets, while it is more rational in developed markets.

Third, we also observe that both forward- and backward-looking Phillips Curve fails to hold during any time periods in Brazil, Greece, Indonesia, Mexico, South Africa, Romania, and Turkey. This rather interesting finding might be due to a number of reasons. These countries have a long history of high and volatile inflation which might be a reason behind the failure of the Phillips relationship. This is because a period of high and volatile inflation discourages firms from investing as firms are less certain about profitability of investment. In addition, these countries also experienced lower economic growth which makes them very prone to recession and causes higher volatility in their labour markets. It is obvious that further studies should be carried out to determine the exact reasons behind the failure of the Phillips relationship for these countries.

## Notes

(1) This paper is generated from the author's doctoral thesis submitted to Social Science Institute, Department of Economics, Istanbul University, Turkey.

(2) It is also known as Neo-Classical Phillips Curve.

## Appendix A

**Table A1.** *Country Codes and Number of Recessions Different Time Periods*

| Country Name | Code | 1980-1990 | 1990-2000 | 1980-2016 | 1990-2016 | 2000-2016 |
|---|---|---|---|---|---|---|
| | | *40 quarters* | *40 quarters* | *145 quarters* | *105 quarters* | *65 quarters* |
| Argentina | AG | 22 | 17 | 55 | 33 | 16 |
| Australia | AU | 9 | 3 | 15 | 6 | 3 |
| Germany | BD | 13 | 13 | 42 | 29 | 16 |
| Belgium | BG | 6 | 6 | 23 | 17 | 11 |
| Brazil | BR | 19 | 16 | 51 | 32 | 16 |
| Canada | CH | 8 | 3 | 11 | 3 | 0 |
| Chile | CL | 10 | 8 | 32 | 22 | 14 |
| China | CN | 11 | 4 | 23 | 12 | 8 |
| Czech Republic | CZ | - | 13 | 24 | 24 | 11 |
| Denmark | DK | 15 | 11 | 50 | 35 | 24 |
| Spain | ES | 9 | 6 | 32 | 23 | 17 |
| Finland | FN | 5 | 15 | 43 | 38 | 23 |
| France | FR | 2 | 5 | 22 | 20 | 15 |
| Greece | GR | 21 | 14 | 75 | 54 | 40 |
| Hungary | HN | - | 17 | 27 | 27 | 10 |
| Indonesia | ID | 9 | 7 | 17 | 8 | 1 |
| India | IN | 10 | 9 | 24 | 14 | 5 |
| Ireland | IR | 15 | 11 | 47 | 32 | 21 |
| Italy | IT | 7 | 13 | 47 | 40 | 27 |
| Japan | JP | 6 | 16 | 46 | 40 | 24 |
| South Korea | KO | 4 | 4 | 11 | 7 | 3 |
| Mexico | MX | 16 | 4 | 31 | 15 | 11 |
| Malaysia | MY | 3 | 3 | 13 | 10 | 7 |
| Netherlands | NL | 11 | 3 | 31 | 20 | 17 |
| Norway | NW | 12 | 12 | 44 | 32 | 20 |
| Austria | OE | 10 | 2 | 32 | 22 | 20 |
| Philippines | PH | 11 | 7 | 21 | 10 | 3 |
| Poland | PO | - | 7 | 15 | 15 | 8 |
| Portugal | PT | 4 | 7 | 37 | 33 | 26 |
| Romania | RM | 19 | 23 | 56 | 37 | 14 |
| Russia | RS | - | 26 | 37 | 37 | 11 |
| South Africa | SA | 11 | 12 | 28 | 17 | 5 |
| Sweden | SD | 9 | 11 | 32 | 23 | 12 |
| Singapore | SP | 3 | 5 | 24 | 21 | 16 |
| Switzerland | SW | 6 | 13 | 31 | 25 | 12 |
| Thailand | TH | 10 | 8 | 29 | 19 | 11 |
| Turkey | TK | 19 | 18 | 47 | 28 | 10 |
| Taiwan | TW | 6 | 4 | 30 | 24 | 20 |
| United Kingdom | UK | 5 | 6 | 18 | 13 | 7 |
| United States | US | 6 | 2 | 18 | 12 | 10 |
| Venezuela | VE | 20 | 13 | 53 | 33 | 20 |

**Notes:** Numbers in the table show the quarter numbers with negative GDP growth (recession). The "-" denote missing data.

**Table A2.** *Results of Unit Root Tests for Series of backward- and forward-looking EAPC*

<table>
<tr><th></th><th colspan="3">ADF (intercept)</th><th></th><th colspan="3">PP (intercept)</th></tr>
<tr><th></th><th>CPI</th><th>EI</th><th>U_U'</th><th></th><th>CPI</th><th>EI</th><th>U_U'</th></tr>
<tr><td>AG</td><td>0.0791 (L:2|N:126)</td><td>0.0508 (L:2|N:126)</td><td>0.0001 (L:0|N:144)</td><td></td><td>0.0000 (B:7|N:128)</td><td>0.0000 (B:7|N:128)</td><td>0.0000 (B:2|N:144)</td></tr>
<tr><td>AU</td><td>0.0008 (L:1|N:143)</td><td>0.0000 (L:0|N:144)</td><td>0.0010 (L:2|N:142)</td><td></td><td>0.0000 (B:8|N:144)</td><td>0.0000 (B:7|N:144)</td><td>0.0423 (B:6|N:144)</td></tr>
<tr><td>BD</td><td>0.1003 (L:3|N:141)</td><td>0.0355 (L:3|N:141)</td><td>0.0419 (L:4|N:140)</td><td></td><td>0.0000 (B:10|N:144)</td><td>0.0000 (B:10|N:144)</td><td>0.0008 (B:9|N:144)</td></tr>
<tr><td>BG</td><td>0.0000 (L:0|N:144)</td><td>0.0350 (L:1|N:143)</td><td>0.4779 (L:3|N:141)</td><td></td><td>0.0000 (B:9|N:144)</td><td>0.0001 (B:9|N:144)</td><td>0.0718 (B:3|N:144)</td></tr>
<tr><td>BR</td><td>0.2191 (L:2|N:103)</td><td>0.115 (L:3|N:103)</td><td>0.0000 (L:8|N:136)</td><td></td><td>0.0993 (B:1|N:105)</td><td>0.2073 (B:6|N:106)</td><td>0.0000 (B:8|N:144)</td></tr>
<tr><td>CH</td><td>0.0045 (L:4|N:140)</td><td>0.0040 (L:4|N:140)</td><td>0.0000 (L:1|N:143)</td><td></td><td>0.0000 (B:10|N:144)</td><td>0.0000 (B:10|N:144)</td><td>0.0172 (B:5|N:144)</td></tr>
<tr><td>CL</td><td>0.7175 (L:7|N:137)</td><td>0.7916 (L:7|N:137)</td><td>0.0002 (L:1|N:119)</td><td></td><td>0.0000 (B:8|N:144)</td><td>0.0001 (B:7|N:144)</td><td>0.0007 (B:4|N:120)</td></tr>
<tr><td>CN</td><td>0.0173 (L:3|N:141)</td><td>0.0106 (L:2|N:142)</td><td>0.0431 (L:0|N:144)</td><td></td><td>0.0000 (B:8|N:144)</td><td>0.0084 (B:4|N:144)</td><td>0.0241 (B:4|N:144)</td></tr>
<tr><td>CZ</td><td>0.0677 (L:3|N:96)</td><td>0.0002 (L:0|N:100)</td><td>0.0081 (L:5|N:87)</td><td></td><td>0.0000 (B:3|N:99)</td><td>0.0000 (B:17|N:100)</td><td>0.0509 (B:3|N:92)</td></tr>
<tr><td>DK</td><td>0.0361 (L:4|N:140)</td><td>0.0273 (L:3|N:141)</td><td>0.5460 (L:1|N:143)</td><td></td><td>0.0000 (B:10|N:144)</td><td>0.0000 (B:9|N:144)</td><td>0.2476 (B:7|N:144)</td></tr>
<tr><td>ES</td><td>0.3683 (L:7|N:137)</td><td>0.0393 (L:7|N:137)</td><td>0.6306 (L:1|N:143)</td><td></td><td>0.0000 (B:8|N:144)</td><td>0.0003 (B:10|N:144)</td><td>0.3053 (B:7|N:144)</td></tr>
<tr><td>FN</td><td>0.0154 (L:4|N:140)</td><td>0.0120 (L:4|N:140)</td><td>0.0001 (L:4|N:140)</td><td></td><td>0.0000 (B:10|N:144)</td><td>0.0001 (B:10|N:144)</td><td>0.0282 (B:9|N:144)</td></tr>
<tr><td>FR</td><td>0.0308 (L:11|N:133)</td><td>0.0239 (L:0|N:144)</td><td>0.0159 (L:1|N:143)</td><td></td><td>0.0055 (B:9|N:144)</td><td>0.0298 (B:12|N:144)</td><td>0.0371 (B:4|N:144)</td></tr>
<tr><td>GR</td><td>0.3312 (L:4|N:140)</td><td>0.5571 (L:4|N:140)</td><td>0.0000 (L:8|N:136)</td><td></td><td>0.0000 (B:10|N:144)</td><td>0.0000 (B:11|N:144)</td><td>0.036 (B:7|N:144)</td></tr>
<tr><td>HN</td><td>0.3649 (L:3|N:141)</td><td>0.4055 (L:6|N:138)</td><td>0.0098 (L:1|N:99)</td><td></td><td>0.0000 (B:10|N:144)</td><td>0.0000 (B:11|N:144)</td><td>0.0714 (B:0|N:100)</td></tr>
<tr><td>ID</td><td>0.0000 (L:0|N:144)</td><td>0.0000 (L:1|N:143)</td><td>0.0000 (L:4|N:140)</td><td></td><td>0.0000 (B:1|N:144)</td><td>0.0018 (B:9|N:144)</td><td>0.0015 (B:2|N:144)</td></tr>
<tr><td>IN</td><td>0.0031 (L:3|N:141)</td><td>0.0009 (L:4|N:140)</td><td>0.0000 (L:4|N:140)</td><td></td><td>0.0000 (B:9|N:144)</td><td>0.0000 (B:10|N:144)</td><td>0.0000 (B:10|N:144)</td></tr>
<tr><td>IR</td><td>0.0067 (L:4|N:140)</td><td>0.0011 (L:4|N:140)</td><td>0.1113 (L:2|N:142)</td><td></td><td>0.0000 (B:7|N:144)</td><td>0.0000 (B:3|N:144)</td><td>0.0652 (B:8|N:144)</td></tr>
<tr><td>IT</td><td>0.0061 (L:8|N:136)</td><td>0.0001 (L:9|N:135)</td><td>0.0024 (L:0|N:144)</td><td></td><td>0.0673 (B:9|N:144)</td><td>0.0881 (B:10|N:144)</td><td>0.0027 (B:6|N:144)</td></tr>
<tr><td>JP</td><td>0.0001 (L:1|N:143)</td><td>0.0086 (L:2|N:142)</td><td>0.6762 (L:0|N:144)</td><td></td><td>0.0000 (B:9|N:144)</td><td>0.0000 (B:8|N:144)</td><td>0.3394 (B:6|N:144)</td></tr>
<tr><td>KO</td><td>0.0001 (L:3|N:141)</td><td>0.0149 (L:2|N:142)</td><td>0.1671 (L:2|N:142)</td><td></td><td>0.0000 (B:8|N:144)</td><td>0.0000 (B:9|N:144)</td><td>0.0440 (B:5|N:144)</td></tr>
<tr><td>MX</td><td>0.0415 (L:0|N:144)</td><td>0.3655 (L:9|N:135)</td><td>0.0414 (L:4|N:140)</td><td></td><td>0.0672 (B:7|N:144)</td><td>0.0901 (B:6|N:144)</td><td>0.0567 (B:6|N:144)</td></tr>
<tr><td>MY</td><td>0.0000 (L:0|N:144)</td><td>0.0027 (L:1|N:143)</td><td>0.0006 (L:0|N:124)</td><td></td><td>0.0000 (B:4|N:144)</td><td>0.0000 (B:8|N:144)</td><td>0.001 (B:4|N:124)</td></tr>
<tr><td>NL</td><td>0.0274 (L:3|N:141)</td><td>0.0040 (L:4|N:140)</td><td>0.1580 (L:12|N:132)</td><td></td><td>0.0000 (B:10|N:144)</td><td>0.0000 (B:10|N:144)</td><td>0.0557 (B:5|N:144)</td></tr>
<tr><td>NW</td><td>0.0602 (L:3|N:141)</td><td>0.0000 (L:3|N:141)</td><td>0.2180 (L:0|N:144)</td><td></td><td>0.0000 (B:9|N:144)</td><td>0.0000 (B:10|N:144)</td><td>0.0848 (B:5|N:144)</td></tr>
<tr><td>OE</td><td>0.0059 (L:4|N:140)</td><td>0.0101 (L:3|N:141)</td><td>0.0842 (L:0|N:144)</td><td></td><td>0.0000 (B:9|N:144)</td><td>0.0000 (B:9|N:144)</td><td>0.1008 (B:1|N:144)</td></tr>
<tr><td>PH</td><td>0.0002 (L:2|N:142)</td><td>0.0005 (L:2|N:142)</td><td>0.0001 (L:4|N:140)</td><td></td><td>0.0000 (B:7|N:144)</td><td>0.0000 (B:7|N:144)</td><td>0.0000 (B:9|N:144)</td></tr>
<tr><td>PO</td><td>0.0964 (L:9|N:108)</td><td>0.2312 (L:6|N:112)</td><td>0.0066 (L:2|N:106)</td><td></td><td>0.0001 (B:3|N:117)</td><td>0.0076 (B:3|N:118)</td><td>0.0768 (B:6|N:108)</td></tr>
<tr><td>PT</td><td>0.5028</td><td>0.3723</td><td>0.1767</td><td></td><td>0.0000</td><td>0.0005</td><td>0.0000 (B:9|N:144)</td></tr>
</table>

<table>
<tr><th></th><th colspan="3">ADF (intercept)</th><th></th><th colspan="3">PP (intercept)</th></tr>
<tr><th></th><th>CPI</th><th>EI</th><th>U_U'</th><th></th><th>CPI</th><th>EI</th><th>U_U'</th></tr>
<tr><td></td><td>(L:7|N:137)</td><td>(L:7|N:137)</td><td>(L:1|N:143)</td><td></td><td>(B:10|N:144)</td><td>(B:10|N:144)</td><td></td></tr>
<tr><td>RM</td><td>0.0000 (L:0|N:101)</td><td>0.0000<br>(L:0|N:102)</td><td>0.0001<br>(L:4|N:108)</td><td></td><td>0.0000 (B:8|N:101)</td><td>0.0000 (B:8|N:102)</td><td>0.0022<br>(B:7|N:112)</td></tr>
<tr><td>RS</td><td>0.1050 (L:2|N:97)</td><td>0.1158<br>(L:1|N:99)</td><td>0.0004<br>(L:4|N:100)</td><td></td><td>0.0002 (B:3|N:99)</td><td>0.0410<br>(B:2|N:100)</td><td>0.0006<br>(B:7|N:104)</td></tr>
<tr><td>SA</td><td>0.0364<br>(L:2|N:142)</td><td>0.2099<br>(L:2|N:142)</td><td>0.0000 (L:5|N:139)</td><td></td><td>0.0000 (B:7|N:144)</td><td>0.0007<br>(B:9|N:144)</td><td>0.0000<br>(B:10|N:144)</td></tr>
<tr><td>SD</td><td>0.1198<br>(L:3|N:141)</td><td>0.0463<br>(L:3|N:141)</td><td>0.1913<br>(L:1|N:143)</td><td></td><td>0.0000 (B:9|N:144)</td><td>0.0002<br>(B:9|N:144)</td><td>0.1991<br>(B:7|N:144)</td></tr>
<tr><td>SP</td><td>0.0000 (L:0|N:144)</td><td>0.0035<br>(L:3|N:141)</td><td>0.0000 (L:1|N:143)</td><td></td><td>0.0000 (B:4|N:144)</td><td>0.0000 (B:3|N:144)</td><td>0.0159<br>(B:10|N:144)</td></tr>
<tr><td>SW</td><td>0.0496<br>(L:4|N:140)</td><td>0.1493<br>(L:3|N:141)</td><td>0.0000 (L:1|N:143)</td><td></td><td>0.0000<br>(B:10|N:144)</td><td>0.0000 (B:9|N:144)</td><td>0.0228<br>(B:7|N:144)</td></tr>
<tr><td>TH</td><td>0.0000 (L:0|N:144)</td><td>0.0003<br>(L:1|N:143)</td><td>0.0000 (L:4|N:140)</td><td></td><td>0.0000 (B:7|N:144)</td><td>0.0000 (B:6|N:144)</td><td>0.0002<br>(B:1|N:144)</td></tr>
<tr><td>TK</td><td>0.5735<br>(L:3|N:141)</td><td>0.6090<br>(L:2|N:142)</td><td>0.0000 (L:8|N:136)</td><td></td><td>0.0000 (B:9|N:144)</td><td>0.0099<br>(B:7|N:144)</td><td>0.0000<br>(B:10|N:144)</td></tr>
<tr><td>TW</td><td>0.0000 (L:3|N:141)</td><td>0.0042<br>(L:6|N:138)</td><td>0.1369<br>(L:5|N:139)</td><td></td><td>0.0000 (B:7|N:144)</td><td>0.0000 (B:6|N:144)</td><td>0.1121<br>(B:9|N:144)</td></tr>
<tr><td>UK</td><td>0.0117<br>(L:4|N:140)</td><td>0.0089<br>(L:4|N:140)</td><td>0.0413<br>(L:2|N:142)</td><td></td><td>0.0000<br>(B:10|N:144)</td><td>0.0000 (B:9|N:144)</td><td>0.0064<br>(B:8|N:144)</td></tr>
<tr><td>US</td><td>0.0002<br>(L:2|N:142)</td><td>0.0000<br>(L:4|N:140)</td><td>0.0295<br>(L:5|N:139)</td><td></td><td>0.0000 (B:7|N:144)</td><td>0.0000 (B:3|N:144)</td><td>0.0022<br>(B:6|N:144)</td></tr>
<tr><td>VE</td><td>0.0074<br>(L:0|N:143)</td><td>0.4144<br>(L:0|N:111)</td><td>0.0000 (L:4|N:140)</td><td></td><td>0.0095<br>(B:8|N:143)</td><td>0.2463<br>(B:2|N:111)</td><td>0.0000 (B:9|N:144)</td></tr>
</table>

**Notes:** Numbers in the table are rejection probabilities of the null hypotheses of ADF and PP tests including intercept. Probabilities below 0.10 denote rejection of these null hypotheses, thus, confirm stationarity of the CPI (inflation), EI (expected inflation), and U_U' (unemployment gap) series of related countries. The lag and observation parameters are presented in the parentheses where “L”, “B”, and “N” denote lag length, Newey-West bandwidth using Bartlett kernel, and observation number respectively. The lag length is determined by Schwarz Information Criterion (SIC) under maximum lag length specification of 13.